%% file: main.tex
\documentclass{Interspeech2024}

\usepackage{cite}
\usepackage{multirow}
\usepackage{multicol}
\usepackage{hyperref}
\usepackage{subcaption}
\usepackage{tabularx}
\usepackage{lipsum}

\interspeechcameraready 

\title{Prompting Large Language Models with Audio for\\ General-Purpose Speech Summarization}

\name[]{Wonjune}{Kang}
\name[]{Deb}{Roy}

\address{
  Massachusetts Institute of Technology
}
\email{wjkang@mit.edu}

\keywords{speech summarization, large language model, speech language model, modality invariance}

\begin{document}

\maketitle


\input{0_abstract}
\input{1_introduction}
\input{2_related_work}
\input{3_method}
\input{4_experimental_setup}
\input{5_results}
\input{6_conclusion}

\clearpage
\bibliographystyle{IEEEtran}
\bibliography{references}

\end{document}

%% file: 0_abstract.tex
\begin{abstract}
In this work, we introduce a framework for speech summarization that leverages the processing and reasoning capabilities of large language models (LLMs).
We propose an end-to-end system that combines an instruction-tuned LLM with an audio encoder that converts speech into token representations that the LLM can interpret.
Using a dataset with paired speech-text data, the overall system is trained to generate consistent responses to prompts with the same semantic information regardless of the input modality.
The resulting framework allows the LLM to process speech inputs in the same way as text, enabling speech summarization by simply prompting the LLM.
Unlike prior approaches, our method is able to summarize spoken content from any arbitrary domain, and it can produce summaries in different styles by varying the LLM prompting strategy.
Experiments demonstrate that our approach outperforms a cascade baseline of speech recognition followed by LLM text processing.
\end{abstract}

%% file: 1_introduction.tex
\section{Introduction}

Large language models (LLMs)~\cite{achiam2023gpt, anil2023palm, touvron2023llama} have demonstrated remarkable capabilities in a wide variety of natural language tasks, such as question answering, machine translation, language understanding, and text summarization.
Notably, they exhibit the ability to respond to prompts of tasks that they were not explicitly trained on without any further gradient updates to their parameters.
These characteristics of LLMs, combined with instruction tuning to align their behavior with human preferences~\cite{wei2022finetuned, ouyang2022training, chung2022scaling}, have allowed users to seamlessly interact with them via text-based prompting.

In this work, we aim to utilize the processing and reasoning capabilities of LLMs for speech-to-text summarization, the task of condensing the linguistic content in a speech signal into a text summary.
We build on recent work that extends the capabilities of LLMs to other modalities, in particular speech and audio.
Speech summarization has traditionally been done using a cascade of automatic speech recognition (ASR) and text summarization~\cite{zhu2020hierarchical, zhang2021exploratory}.
Recently, some works have proposed systems for end-to-end speech summarization~\cite{sharma2022end, matsuura2023leveraging, matsuura23transfer}, which aim to reduce errors caused by the propagation of mistakes in the ASR step.
However, prior methods are limited by their training data; models trained on a specific dataset are restricted to producing summaries in that domain and in the style of that dataset's annotations.
Meanwhile, LLMs are able to summarize text from arbitrary domains and in different styles by varying their prompts; we would like to leverage these abilities to develop a system that can perform general-purpose speech summarization.

To this end, we propose an end-to-end pipeline that combines a text LLM with an audio encoder that allows the LLM to understand and process speech inputs in the same way as text.
Our model is trained using the concept of \textit{modality invariance}---the idea that, given certain semantic information in a prompt, the LLM should provide the same response regardless of the prompt's modality~\cite{fathullah2023towards}.
Specifically, we use an ASR dataset with paired speech-text data; while keeping the LLM weights frozen, we train the audio encoder to convert speech inputs into token representations that the LLM can interpret.
Then, the end-to-end system is guided to produce the same output as when text is the input using next-token prediction loss.
We additionally incorporate knowledge distillation using the response from the corresponding text input as the teacher model, utilizing feature and logit distillation losses to guide the model to produce more consistent responses from speech inputs.

Under this framework, every capability of the LLM with text also becomes possible with speech, and we can perform speech summarization by simply prompting the LLM to do so.
Because our method takes advantage of the innate abilities of LLMs, it can be used for summarizing spoken content or documents from any arbitrary domain.
Furthermore, it is able to generate summaries in different styles by varying the prompting strategy to the LLM.
Experiments demonstrate that our proposed system outperforms a baseline cascade of ASR and LLM-based text processing in terms of response language modeling and speech summarization on the CNN / DailyMail dataset~\cite{nallapati2016abstractive}.

In summary, the contributions of this work are as follows: 1) We propose a framework that allows an LLM to understand and process any speech input, and leverage its capabilities to perform speech summarization. 2) Our method can be used to summarize spoken content from any arbitrary domain. 3) We demonstrate the system's ability to produce summaries in different styles by adjusting the prompt to the LLM.\footnote{Code available at: \url{https://github.com/wonjune-kang/llm-speech-summarization}. Examples of summaries from various prompts and speech utterances are included on our demo page: \url{https://llm-speech-summarization.github.io/}}

%% file: 2_related_work.tex
\section{Background and Related Work}

\subsection{Large language models (LLMs) and speech}
\vspace{-2pt}

There has been a large body of recent work on extending the capabilities of LLMs to speech and audio.
Most approaches utilize some sort of audio encoder to process audio into discrete tokens that can be fed into the LLM as inputs.
These audio tokens can be used to augment the original text-only token vocabulary of the LLM~\cite{rubenstein2023audiopalm, zhang2023speechgpt} or trained in a way that aligns them with the original text token embedding space~\cite{fathullah2024prompting, fathullah2023towards}.
Some models~\cite{rubenstein2023audiopalm, fathullah2024prompting, wu2023decoder} are trained to perform specific tasks such as speech translation, speech recognition, or speech synthesis, while others~\cite{zhang2023speechgpt, tang2024salmonn} are designed to be treated as general purpose models that can process a broad range of speech prompts.

The approach we use in this work is most similar to AudioChatLlama~\cite{fathullah2023towards}, which utilizes the concept of \textit{modality invariance} between prompts.
Given certain semantic information in a prompt, the idea is that an LLM should provide the same response regardless of whether the prompt is in audio or text.
AudioChatLlama consists of a trainable audio encoder and a frozen LLM; using paired speech-text samples, the overall system is guided to produce the same response whether the input to the LLM is in audio or text.
However, it only relies on next-token prediction loss to align its outputs with those generated from text inputs, and the authors did not study any temporal alignment strategies between the two modalities when designing their audio encoder.
In contrast, we also incorporate knowledge distillation by matching the LLM's token prediction logits and hidden states between text and audio inputs, and we explore various design choices for the audio encoder so as to induce the LLM to produce more consistent responses to speech prompts.

\vspace{-2pt}
\subsection{Speech summarization}
\vspace{-2pt}

Recent works on speech summarization have largely focused on end-to-end models~\cite{sharma2022end, matsuura2023leveraging, matsuura23transfer}, which aim to mitigate error propagation from mistakes made during intermediate steps in traditional cascade models of ASR and text summarization~\cite{zhu2020hierarchical, zhang2021exploratory}.
However, like most machine learning models, these previous methods are inherently limited by their training data; models trained on a specific dataset are only able to produce summaries in that domain and in the style of that dataset's annotations.
Recently, \cite{jung2024augsumm} proposed a method to leverage LLMs as a proxy for human annotators to generate a wider variety of summaries for training and evaluating summarization models.
However, the fundamental problem still stands; for example, a model trained to summarize instructions using the How2 dataset~\cite{sanabria2018how2} would likely not be suitable for summarizing the content of long meetings.
In contrast, our approach performs speech summarization by leveraging the innate abilities of LLMs, which enables a more general system that is capable of handling diverse domains and stylistic preferences.

%% file: 3_method.tex
\vspace{-2pt}
\section{Proposed Method}

\subsection{System architecture}
\vspace{-2pt}

Our proposed system consists of two components: an LLM and an audio encoder that translates speech inputs into token embeddings that the LLM can receive as part of a prompt.
We use MiniChat-3B as our LLM, an instruction-tuned version of MiniMA-3B~\cite{zhang2023towards}, which was distilled from Llama 2 7B~\cite{touvron2023llama}.
Although small for LLM standards, it achieves strong performance on a variety of benchmarks compared to other 3B parameter models and approaches the performance of Llama 2 7B, making it a suitable choice for us.
Like most other decoder-only language models, MiniChat takes a token embedding sequence as input and autoregressively generates a text response.

The backbone of the audio encoder is initialized with a HuBERT-Large model~\cite{hsu2021hubert} that was fine-tuned to perform ASR on the Librispeech corpus~\cite{panayotov2015librispeech} using connectionist temporal classification (CTC) loss~\cite{graves2006connectionist}.
It takes as input mono-channel audio at a 16 kHz sampling rate.
Following the HuBERT layers, which produce 1024-dim representations every 20 ms (50 Hz), we perform temporal downsampling by a factor of 4 by average pooling with a kernel size of 8 and a stride of 4.
This is followed by a linear projection layer to match the dimension of the pooled representations with that of the LLM (3072-dim).
The encoder thus produces audio tokens at a rate of 12.5 Hz.

We also explored a downsampling method which involved pooling HuBERT representations based on the predicted CTC alignments of words from a separate HuBERT ASR model.
Specifically, we performed the same average pooling as above, but between the CTC-aligned start embedding indices of consecutive words predicted by the ASR model; this was an effort to more explicitly align audio tokens with words.
We discuss the results of these experiments in Section \ref{sec:results}.

\begin{figure}[t!]
    \centering
    \includegraphics[width=\columnwidth]{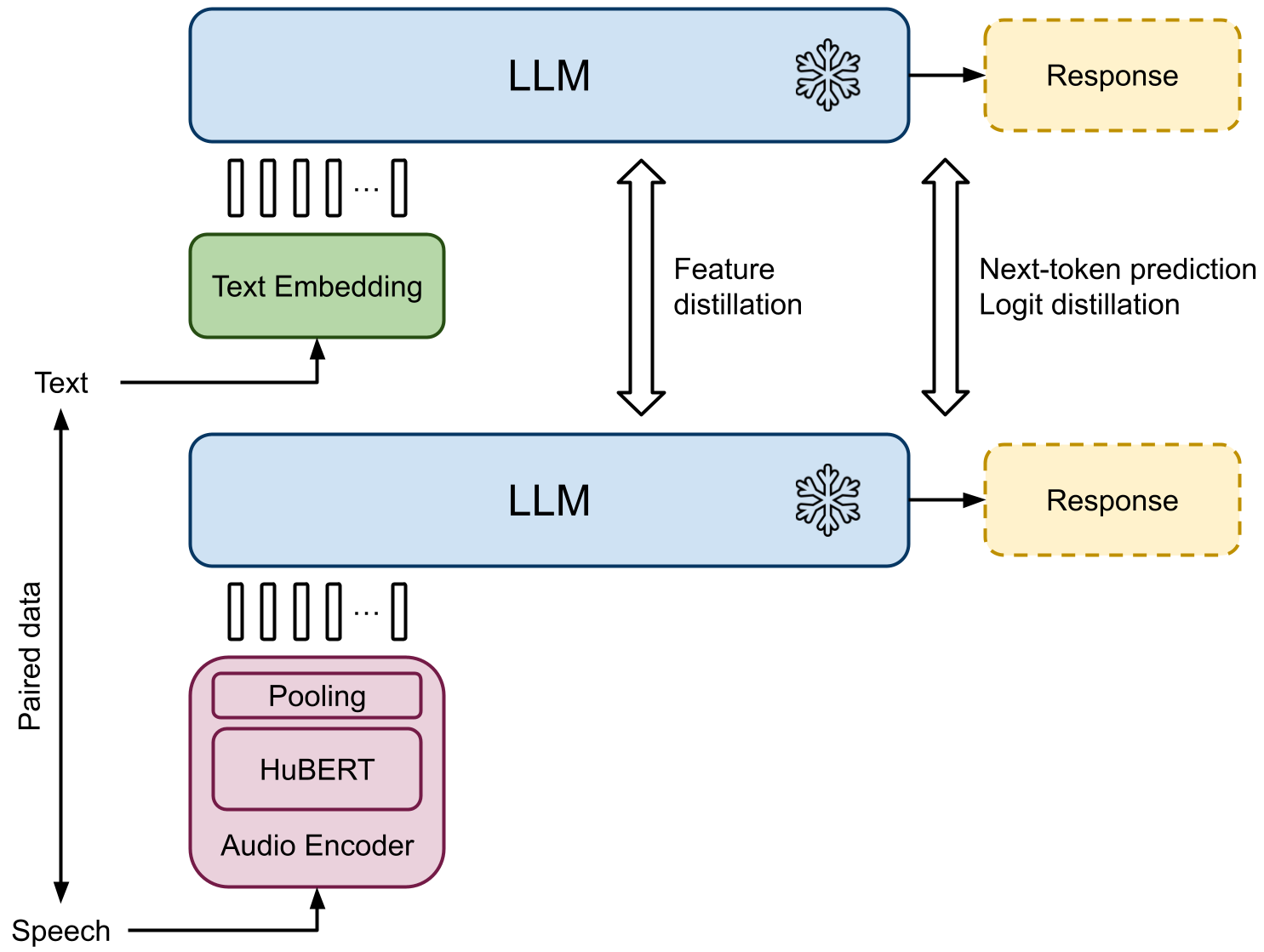}
    \caption{The audio encoder is trained to produce token embeddings such that the LLM's response to a spoken prompt matches its response to a text prompt with the same semantic content.}
    \label{fig:training}
    \vspace{-6pt}
\end{figure}

\vspace{-4pt}
\subsection{Aligning responses for audio and text inputs}
\vspace{-1pt}

Figure \ref{fig:training} shows the training strategy for the audio encoder and LLM pipeline.
At a high level, it involves aligning the LLM's response to an audio prompt with the response to a matching text prompt with the same semantic information.
We would like to utilize the capabilities of our LLM without compromising any of its original performance; therefore, we keep the LLM's weights frozen during all of our experiments, with the audio encoder being the only trainable component of our system.
Although this framework prevents the LLM from utilizing any additional information in the speech input beyond the linguistic content, it allows us to safely preserve its original abilities without risk of overfitting to any specific tasks or datasets.

\vspace{2pt}
\noindent \textbf{Data.}
We make use of the Librispeech corpus~\cite{panayotov2015librispeech}, which is often used for ASR and contains paired audio-text samples of spoken utterances and their transcripts.
For all of the samples in Librispeech, we prompt the LLM using the text transcripts and save its responses to create a dataset with (audio, text, response) components for each sample.
These responses are then treated as the ground truth output which the audio encoder is trained to induce from the LLM.
Note that Librispeech consists of utterances that are collected from audiobooks, which generally do not make for useful prompts when fed into an LLM without any additional context.
However, for our training framework, we can consider the objective to have been achieved if the audio prompt can produce the same response as the text prompt, even if the prompt and LLM response are nonsensical.

\vspace{2pt}
\noindent \textbf{Training.}
We train our system using next-token prediction (NTP) loss, conditioning the prediction of the response tokens on the input audio tokens.
Formally, let the full token sequence fed into the model be $\mathbf{x} = (x_0, ..., x_t, x_{t+1}, ... x_T)$, where the first $t$ elements correspond to the prompt, including the audio tokens and any tokens used in the LLM's chat structure: $(x_0, ..., x_t)$.
The remaining elements correspond to the tokens for the LLM's ground truth response to the corresponding text prompt: $(x_{t+1}, ..., x_T)$.
Then, the NTP loss is:
\begin{align}
    \mathcal{L}_{\mathrm{NTP}} = -\sum_{i = t+1}^{T} \log p(x_i | x_{<i}).
\end{align}
We additionally incorporate knowledge distillation at the logit~\cite{hinton2015distilling} and feature~\cite{heo2019comprehensive} levels using the outputs and activations of the LLM given the original text input as the teacher model.
These losses are meant to more explicitly guide the audio encoder to induce predictions and hidden states from the LLM that match those from the standard text input.
For logit distillation (LD), we perform logit matching between the text and audio prompts using a soft cross-entropy loss as follows:
\begin{align}
    \mathcal{L}_{\mathrm{LD}} = -\sum_{i = t+1}^{T} \sum_{j=1}^{|\mathcal{V}|} \Tilde{y}^{(j)}_i \log \hat{y}^{(j)}_i,
\end{align}
where $\mathcal{V}$ is the SentencePiece token vocabulary of the LLM, and $\Tilde{y}$ and $\hat{y}$ denote the next-token prediction logits conditioned on the text and audio prompts, respectively.
For feature distillation (FD), we use the mean-square error (MSE) loss between the hidden states of the LLM when given the text prompt ($\mathbf{\Tilde{h}}$) and audio prompt ($\mathbf{\hat{h}}$) at connector layers $L$ as follows:
\begin{align}
    \mathcal{L}_{\mathrm{FD}} = \sum_{l \in L} \mathrm{MSE} \big( \mathbf{\Tilde{h}}^{l}_{t+1:T}, \mathbf{\hat{h}}^{l}_{t+1:T} \big).
\end{align}
The overall training objective is thus:
\begin{align}
    \mathcal{L}_{\mathrm{total}} = \lambda_{\mathrm{NTP}} \mathcal{L}_{\mathrm{NTP}} + \lambda_{\mathrm{LD}} \mathcal{L}_{\mathrm{LD}} + \lambda_{\mathrm{FD}} \mathcal{L}_{\mathrm{FD}}
\end{align}
with weighting factors $\lambda_{\mathrm{NTP}}$, $\lambda_{\mathrm{LD}}$, and $\lambda_{\mathrm{FD}}$ for each loss.
Note that all losses are computed only for token indices corresponding to the response portion of the sequence, $[t+1, T]$.

\begin{figure}[t!]
    \centering
    \includegraphics[width=\columnwidth]{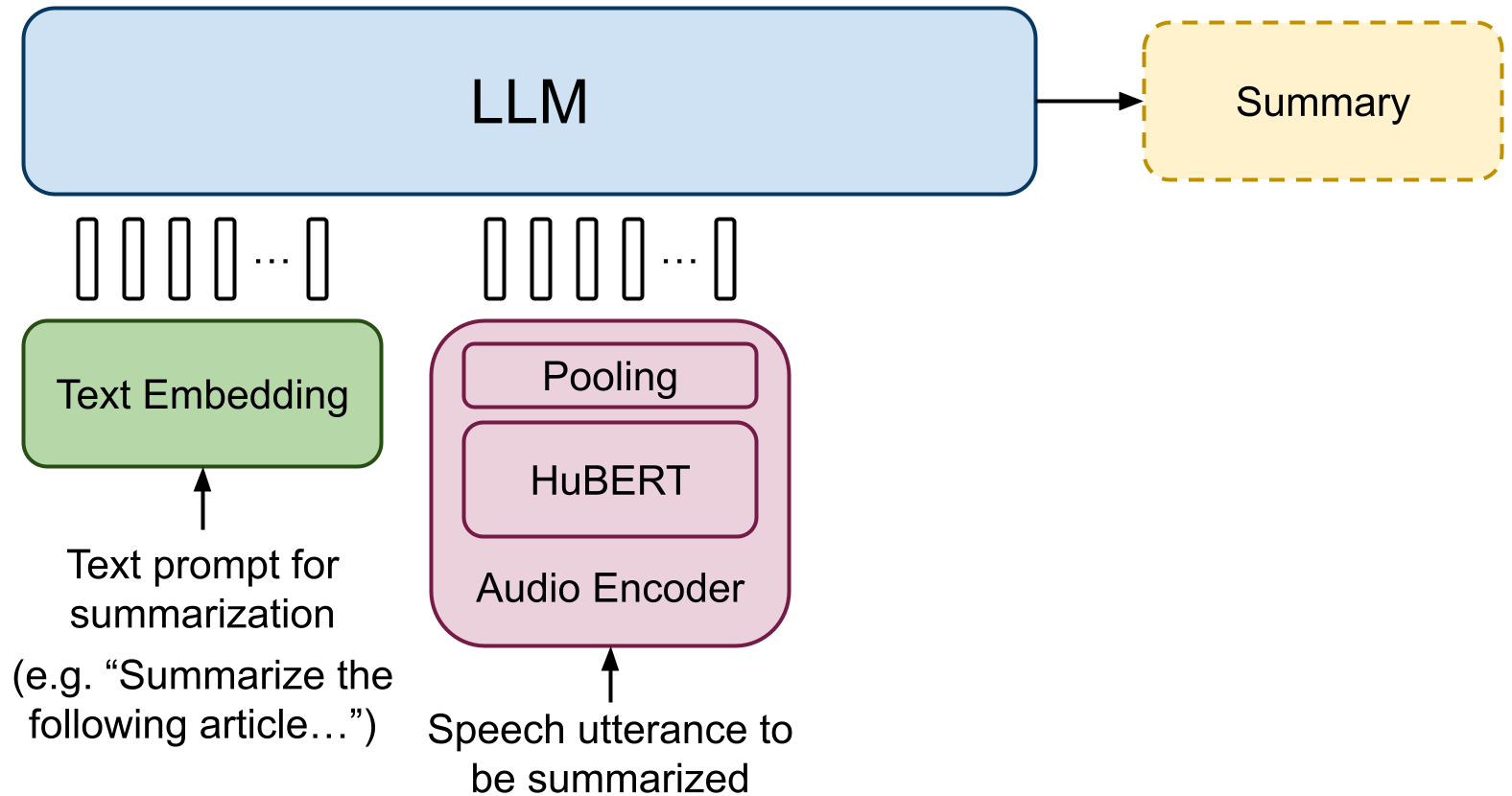}
    \caption{Speech summarization can be performed by prompting the LLM using any combination of text and speech inputs.}
    \label{fig:inference}
    \vspace{-6pt}
\end{figure}

\vspace{-3pt}
\subsection{Performing speech summarization}
\vspace{-1pt}

Once the pipeline consisting of the audio encoder and LLM has been trained, the LLM can take in any combination of text and speech inputs as part of a user's prompt.
Speech summarization can thus be performed by simply instructing the LLM to do so in the prompt; Figure \ref{fig:inference} shows the inference process.
Because our system leverages the general-purpose abilities of LLMs to process any arbitrary input, it can summarize spoken content from any domain.
We can further control the style of the desired summary by adjusting the prompt, such as by instructing the LLM to vary the summary's length or to emphasize or de-emphasize specific aspects of the content.

%% file: 4_experimental_setup.tex
\begin{table}[t]
    \centering
    \caption{Perplexity of the LLM on the Librispeech test sets under the ``correct'’ response.}
    \vspace{-6pt}
    \footnotesize
    \begin{tabular}{lcc}
    \toprule
    \multirow{2}{*}{\vspace{-5pt} \textbf{Model}} &
    \multicolumn{2}{c}{\textbf{Perplexity}}  \\
    \cmidrule(lr){2-3}
    & \multicolumn{1}{l}{\textit{test-clean}}
    & \multicolumn{1}{l}{\textit{test-other}} \\ \midrule
    Ground truth text + LLM     & 1.608 & 1.595 \\ \midrule
    Cascade: HuBERT ASR + LLM   & 1.644 & 1.682 \\ \midrule
    Proposed system             & 1.604 & 1.638 \\
    \quad NTP + FD losses only  & 1.552 & 1.589 \\
    \quad NTP + LD losses only  & 1.592 & 1.627 \\
    \quad NTP loss only         & 1.531 & 1.568 \\
    \quad CTC-aligned pooling   & 1.603 & 1.637 \\ \bottomrule
    \end{tabular}
    \vspace{-10pt}
    \label{table:ppl_results}
\end{table}

\begin{table*}[t]
    \centering
    \caption{Speech summarization performance on the curated CNN / DailyMail test set with speech synthesized using TTS. Metrics are measured against the original summaries from the CNN / DailyMail dataset as well as synthetic ones generated by Llama 2-Chat 7B.}
    \vspace{-6pt}
    \footnotesize
    \begin{tabular}{lcccccc}
    \toprule
    \multirow{2}{*}{\vspace{-5.5pt} \textbf{Model}} &
    \multicolumn{3}{c}{\textbf{CNN / DailyMail Summaries}} &
    \multicolumn{3}{c}{\textbf{Llama 2-Chat 7B Summaries}} \\
    \cmidrule(lr){2-4}
    \cmidrule(lr){5-7}
    &
    ROUGE-1, 2, L &
    METEOR &
    BERTScore &
    ROUGE-1, 2, L &
    METEOR &
    BERTScore \\ \midrule
    Ground truth text + LLM        & 37.31, 15.69, 24.66                   & 38.80             & 87.72             & 52.94, 28.49, 36.76                   & 44.64             & 91.08 \\ \midrule
    Cascade: HuBERT ASR + LLM   & \textbf{33.60}, 12.97, \textbf{22.56} & 34.73             & \textbf{86.87}    & 49.18, 25.56, 34.86                   & 41.20             & 90.13 \\ \midrule
    Proposed system             & 33.34, \textbf{13.21}, 22.52          & \textbf{36.46}    & 86.83             & 51.12, \textbf{27.50}, \textbf{37.48} & \textbf{45.63}    & \textbf{90.50} \\
    \quad NTP + FD losses only  & 33.03, 12.81, 22.24                   & 36.20             & 86.74             & 51.33, 26.70, 36.74                   & 44.85             & 90.28 \\
    \quad NTP + LD losses only  & 32.37, 12.51, 21.65                   & 35.88             & 86.63             & 51.26, 26.93, 36.37                   & 45.40             & 90.29 \\
    \quad NTP loss only         & 31.66, 12.20, 21.03                   & 36.04             & 86.51             & 50.47, 26.01, 35.50                   & 45.62             & 90.06 \\
    \quad CTC-aligned pooling   & 32.53, 12.36, 21.70                   & 35.83             & 86.70             & \textbf{51.60}, 27.05, 36.89          & 45.51             & 90.39 \\ \bottomrule
    \end{tabular}
    \label{table:summarization_results}
    \vspace{-8pt}
\end{table*}

\section{Experimental Setup}
\vspace{-3pt}

\subsection{Datasets}
\vspace{-3pt}

We trained our system using the Librispeech corpus~\cite{panayotov2015librispeech}, using the standard splits for training (960 hours), validation, and testing.
To create the LLM response dataset, we fed reference text into MiniChat and generated the response using greedy decoding with a maximum decoding length equal to twice the length of the input prompt.

To evaluate speech summarization, we used the CNN / DailyMail dataset~\cite{nallapati2016abstractive}.
Because it does not include audio, we synthesized speech for the dataset using a text-to-speech (TTS) model, StyleTTS 2~\cite{li2023styletts} trained on the LJ Speech dataset~\cite{ljspeech17}.
We only used the test set, and to keep audio file lengths manageable, we selected articles that were no more than 1600 characters in length.
This resulted in 886 samples where synthesized speech utterances ranged from 21.3 to 149.3 seconds in length, with an average of 93.9 seconds.
While synthesized speech is preferably avoided, there are unfortunately few speech to summary datasets available.
We opted not to use the popular How2 dataset~\cite{sanabria2018how2} because we were interested in evaluating summarization in a domain beyond what that dataset specifically covers (instructional videos).
Furthermore, How2 only includes pre-extracted acoustic features and not raw waveforms, and thus would also have required some method to re-synthesize audio.

Additionally, we found the ground truth annotations in CNN / DailyMail to be ill-suited for evaluating LLM-generated summaries.
They consist of concatenated bullet points that are not continuous pieces of text, and are thus quite different from the text style produced by LLMs.
Because we were interested in performing and evaluating LLM-style summarization, we created a new set of reference summaries by prompting Llama 2-Chat 7B~\cite{touvron2023llama} to summarize the CNN / DailyMail articles in 3 sentences or less.
We evaluated performance against both the original summaries and the Llama 2-generated ones.

\vspace{-4pt}
\subsection{Training and evaluation}
\vspace{-2pt}

We conducted all experiments on a single NVIDIA V100 32GB GPU.
All models were trained for up to 2.4M steps with a batch size of 1 and gradient accumulation every 16 steps (effectively 150k steps with a batch size of 16).
We used the AdamW optimizer~\cite{loshchilov2018decoupled} with $\beta_1 = 0.9$, $\beta_2 = 0.999$, with an initial learning rate of 5e-5 that linearly decayed down to 5e-6 over the course of training.
We empirically set $\lambda_\mathrm{NTP} = 0.5$, $\lambda_\mathrm{LD} = 0.5$, and $\lambda_\mathrm{FD} = 1.0$.
For feature distillation, we used connector layers at indices $L = [1, 6, 12, 18, 24]$ of the 24-layer LLM.

We selected the checkpoint with the best validation NTP loss for each model.
When performing summarization, we always prompted the LLM to do so in 3 sentences or less and used greedy decoding for consistency.
Performance was evaluated using ROUGE~\cite{lin2004rouge}, METEOR~\cite{denkowski2014meteor}, and BERTScore~\cite{zhang2019bertscore}.

%% file: 5_results.tex
\section{Results}
\label{sec:results}
\vspace{-1pt}

We compare our proposed system against a baseline cascade system of an ASR model and MiniChat-3B.
For a fair comparison, we use the same HuBERT-Large model as the one used to initialize the backbone of our audio encoder to perform ASR; it achieves 1.9\% and 3.3\% word error rates on the \textit{test-clean} and \textit{test-other} splits of Librispeech, respectively.
We also perform comparisons against ablations of our model with different combinations of training objectives and architectural choices, and include scores from feeding the ground truth text into the LLM as an additional point of reference.

\vspace{-4pt}
\subsection{Response language modeling}
\label{subsec:response_lm}
\vspace{-2.5pt}

As a preliminary experiment, we looked at the LLM's performance on response language modeling by measuring its perplexity under the ``correct'' response to prompts on the test sets of Librispeech; i.e., the response generated when prompted with the ground truth transcript of the audio.
The results are shown in Table~\ref{table:ppl_results}.
All variants of our end-to-end system obtained lower perplexity than the cascade baseline and comparable values to when the ground truth text was fed into the LLM.
Note that perplexity is an imperfect metric for measuring language modeling performance~\cite{meister2021language}, and lower values do not necessarily indicate better performance on downstream tasks.
However, we treat these results as a useful sanity check to demonstrate that our system is able to process speech inputs coherently.

\vspace{-4pt}
\subsection{Speech summarization}
\label{subsec:summarization}
\vspace{-2.5pt}

Table \ref{table:summarization_results} shows the summarization metrics of the various models on the curated CNN / DailyMail test set, measured against both the original reference summaries and the ones generated using Llama 2-Chat.
When evaluating against the original summaries, our proposed method obtained similar scores to the cascade system, and all systems obtained quite low scores overall.
This indicates a fundamental mismatch between the text style of the original annotations and that of the summaries generated by the LLM.
However, all models obtained significantly higher scores when evaluating against summaries generated by Llama 2-Chat, indicating a closer stylistic match to text generated by another instruction-tuned LLM.
Here, our proposed end-to-end system significantly outperformed the cascade baseline in terms of all metrics, and it approached the scores obtained from feeding the actual reference text into the LLM.

When looking at the results of ablations, we found that each of the loss components used for training the audio encoder contributed to performance in a meaningful way, with the system trained on all three losses achieving the best overall performance.
Interestingly, using CTC-based word offsets as the basis to pool embeddings resulted in a slight decrease in scores compared to standard pooling.
This could be a result of using the predicted embedding offsets as ``hard'' alignments for pooling; if the words predicted by the ASR model are incorrect, aggregating audio tokens based on these incorrect offsets could result in a noisier token sequence being provided to the LLM.
The issue could potentially be rectified by using a probabilistic ``soft'' alignment instead, or by using a separate module for speech-token alignment such as the Montreal Forced Aligner~\cite{mcauliffe2017montreal}.

We also explored varying the prompt to the LLM to generate summaries from speech in different styles; Table \ref{table:example_summaries} shows examples of various summaries generated in this way.
These results demonstrate that our framework successfully leverages the reasoning and processing abilities of LLMs, allowing it to be used as a general-purpose speech summarization system that can be applied to any domain or desired summarization style.
Additional examples are included on our demo page.

\begin{table}[t]
    \centering
    \caption{Examples of summaries generated from speech in different styles by varying the user prompt. All prompts begin with: ``Summarize the following article in 3 sentences or less...''}
    \vspace{-6pt}
    \scriptsize
    \begin{tabularx}{\columnwidth}{
        >{\hsize=.27\hsize \raggedright\arraybackslash}X
        >{\raggedright\arraybackslash}X
    }
    \toprule
    \multicolumn{1}{c}{\textbf{Prompt}} & \multicolumn{1}{c}{\textbf{Summary}} \\ \midrule
    \vspace{6pt} no other condition & Juventus is close to winning a fourth straight Serie A title by defeating Lazio 2-0. Carlos Tevez scored his 26th goal of the season, helping Juventus take an early lead. Lazio attempted a late comeback but were unable to score, ending their six-game winning streak. \\ \midrule
    \vspace{2.3pt} ``without mentioning any names'' & Juventus's 2-0 victory over Lazio moved them closer to their fourth consecutive Serie A title. Lazio's 15-point lead over second place was reduced to 19 points after the match. Sampdoria drew 0-0 against Cesena, leaving both teams level on 50 points. \\ \midrule
    \vspace{9.5pt} ``focusing on Lazio''  & Carlos Tevez scored twice as Juventus defeated Lazio 2-0, moving 15 points ahead of second-place rivals. Lazio's six-game winning streak ended, and the team could be dragged back into third place if their city rivals, Roma, beat Atalanta. Sampdoria drew 0-0 against Cesena, leaving both teams level on 50 points. \\ \bottomrule
    \end{tabularx}
    \vspace{-9pt}
    \label{table:example_summaries}
\end{table}

%% file: 6_conclusion.tex
\vspace{-4pt}
\section{Conclusion}
\vspace{-1pt}

We introduced an end-to-end framework for speech-to-text summarization that can be applied towards spoken content from any domain.
The proposed system combines an instruction-tuned LLM with an audio encoder that converts speech into token embeddings that the LLM can interpret, enabling speech summarization by simply prompting the LLM.
Unlike prior methods, our approach is able to generate summaries in different styles by varying the prompts to the LLM.
The quality of the system's components are a key factor for its performance; while we performed experiments using relatively small models in this work, our framework is applicable for any combination of audio encoder and LLM architectures.
Using larger, more powerful models would open up possibilities for significantly more performant general-purpose speech summarization systems.